\documentclass[sigconf]{acmart}

\usepackage{latexsym}
\usepackage{graphicx}
\graphicspath{{./images/}}
\usepackage{booktabs} 
\usepackage{color}  
\usepackage{amsmath}  
\usepackage{subcaption}
\usepackage{caption}
\usepackage{tikz}
\usepackage{colortbl} 
\usepackage{framed}
\usepackage{multirow}
\usepackage{multicol}
\usepackage{hyperref}
\usepackage{url}
\usepackage{balance}
\usepackage{verbatim}
\usepackage{cancel}
\usepackage{xspace} 
\usepackage[ruled,vlined]{algorithm2e}
\usepackage{bbold} 

\newcommand{\struct}[1]{\texttt{\small #1}}
\newcommand{\utterance}[1]{\textit{#1}}
\newcommand{\phrase}[1]{\textit{``#1''}}

\newenvironment{Snugshade}[1][236,236,236]{
    \setlength{\itemsep}{0pt}
     \setlength{\parsep}{0pt}
     \setlength{\topsep}{0pt}
     \setlength{\partopsep}{0pt}
     \setlength{\leftmargin}{1.5em}
     \setlength{\labelwidth}{0em}
     \setlength{\labelsep}{0em} 
\setlength{\parskip}{0pt}
    \definecolor{shadecolor}{RGB}{#1}%
    \begin{snugshade}
}{%
    \end{snugshade}%
}

\newcommand{\convex}{\textsc{Convex}\xspace}
\newcommand{\convquestions}{\textsc{ConvQuestions}\xspace}

\acmConference[CIKM'19]{ACM CIKM Conference}{October 2019}{Beijing, China}

\acmArticle{4}
\acmPrice{15.00}

\begin{document}
\fancyhead{}

\title{Look before you Hop: Conversational Question Answering\\
	over Knowledge Graphs Using Judicious Context Expansion}

\author{Philipp Christmann}
\affiliation{%
  \institution{MPI for Informatics, Germany}
}
\email{pchristm@mmci.uni-saarland.de}

\author{Rishiraj Saha Roy}
\affiliation{%
  \institution{MPI for Informatics, Germany}
}
\email{rishiraj@mpi-inf.mpg.de}

\author{Abdalghani Abujabal}
\affiliation{%
  \institution{Amazon Alexa, Germany}
}
\email{abujabaa@amazon.de}

\author{Jyotsna Singh}
\affiliation{%
  \institution{MPI for Informatics, Germany}
}
\email{jsingh@mpi-inf.mpg.de}

\author{Gerhard Weikum}
\affiliation{%
  \institution{MPI for Informatics, Germany}
}
\email{weikum@mpi-inf.mpg.de}

\renewcommand{\shortauthors}{P. Christmann et al.}

\newcommand{\squishlist}{
 \begin{list}{$\bullet$}
  { \setlength{\itemsep}{0pt}
     \setlength{\parsep}{1pt}
     \setlength{\topsep}{1pt}
     \setlength{\partopsep}{0pt}
     \setlength{\leftmargin}{1.5em}
     \setlength{\labelwidth}{1em}
     \setlength{\labelsep}{0.5em} } }

\newcommand{\squishend}{
  \end{list}  }

\begin{abstract}
Fact-centric information needs are rarely one-shot; users typically ask follow-up questions to explore a topic.
In such a conversational setting, 
the user's inputs are often incomplete, with entities or predicates left out,
and ungrammatical phrases.
This poses a huge challenge to 
question answering (QA) systems
that typically rely on 
cues in 
full-fledged interrogative sentences.
As a solution, we develop \convex: 
an
unsupervised method that can
answer incomplete questions over a knowledge graph (KG) 
by maintaining conversation context using entities and predicates
seen so far and automatically inferring missing or ambiguous 
pieces for follow-up questions.
The core of our method is a graph exploration algorithm that
judiciously expands a frontier to find candidate answers for
the current question.
To evaluate \convex, we release \convquestions,
a crowdsourced benchmark with
$11,200$ distinct conversations from five different domains. We show that
\convex: (i) adds conversational support to any stand-alone QA system,
and (ii) outperforms 
state-of-the-art
baselines
and 
question completion strategies.
\end{abstract}

%
%

\ccsdesc[500]{Information systems~Question answering}



\maketitle

\section{Introduction}
\label{sec:introduction}

\subsection{Motivation}
\label{subsec:motivation}


Obtaining direct answers to fact-centric questions
is supported by large knowledge graphs (KGs) such as Wikidata
or industrial KGs (at Google, Microsoft, Baidu, Amazon, etc.),
consisting of semantically organized entities, attributes, and
relations in the form of subject-predicate-object (SPO) triples.
This task of question answering over KGs (KG-QA) has been
intensively researched 
\cite{berant2013semantic,bast2015more,unger2012template,yahya2013robust,
	abujabal2018never,diefenbach2019qanswer,tanon2018demoing,huang2019knowledge}.
%
However, users' information needs are not always expressed in 
well-formed and self-contained questions for one-shot processing.
Quite often,
users issue a series of follow-up questions to explore
a topic~\cite{saha2018complex,guo2018dialog}, analogous to search 
sessions~\cite{ren2018conversational}. 
A major challenge in such {\em conversational QA} settings
is that follow-up questions are often incomplete, with entities
or predicates not spelled out, and use of ungrammatical phrases.
So a large part of the context is unspecified, assuming that the
systems implicitly understand the user's intent from previous interactions. 
Consider
the following conversation as a running example.
A user
asks questions (or utterances) $q^i$ and the system has to generate answers 
$a^i$:

\begin{Snugshade}
	$q^0$: \utterance{Which actor voiced the Unicorn in The Last Unicorn?}
	
	$a^0$: \textit{Mia Farrow}		
	
	$q^1$: \utterance{And Alan Arkin was behind $\ldots$?}	
	
	$a^1$: \textit{Schmendrick}	
	
	$q^2$: \utterance{Who did the score?}
	
	$a^2$: \textit{Jimmy Webb}	

	$q^3$: \utterance{So who performed the songs?}	
	
	$a^3$: \textit{America}	
	
	
	$q^4$: \utterance{Genre of this band's music?}
	
	$a^4:$ \textit{Folk rock, Soft rock}

	$q^5$: \utterance{By the way, who was the director?}	
	
	$a^5$: \textit{Jules Bass}
\end{Snugshade}

Such conversations are characterized by a well-formed and complete initial
question ($q^0$) with incomplete follow-ups ($q^1 - q^5$),
an initial and often central entity of interest (\phrase{The Last Unicorn}), 
slight shifts in
focus (inquiry of the band America's
genre in $q^4$), informal styles ($q^1, q^5$), and a running context
comprised of
entities and predicates in all preceding questions and answers
(not just immediate precedents).
%

\textbf{Limitations of state-of-the-art KG-QA.} 
State-of-the-art
systems~\cite{huang2019knowledge,abujabal2018never,diefenbach2019qanswer,
	luo2018knowledge,tanon2018demoing}
expect well-formed input questions (like $q^0$), complete with cue words
for 
entities (\phrase{Unicorn}), predicates (\phrase{voiced}), and
types (\phrase{actor}), and map them to corresponding KG-items.
A SPARQL query (or
an equivalent logical expression) is generated 
 to retrieve answers.
For example, a Wikidata query for $q^0$
would be:
\struct{SELECT ?x WHERE \{TheLastUnicorn voiceActor ?x . ?x characterRole 
TheUnicorn\}}. 
In our conversational setup, such methods
completely fall apart
due to the incompleteness of follow-up questions, and the ad-hoc ways
in which they are phrased.

The alternative approach of
\textit{question completion}~\cite{kumar2017incomplete} 
aims to create
{syntactically correct} 
full-fledged 
interrogative sentences 
from 
the user's inputs, closing 
the gaps
by learning from
supervision pairs, 
while being agnostic to
the underlying KG.
However, 
this paradigm is bound to be limited and would fail for
ad-hoc styles of user inputs or when training data is too sparse.


\vspace*{-0.5cm}

\subsection{Approach and Contributions}
\label{subsec:app}

Our proposed approach, \convex (\textbf{CONV}ersational KG-QA with
context \textbf{EX}pansion) overcomes these
limitations, based on the following key ideas.
The initial question is used to identify a small subgraph of the KG for
retrieving answers, similar to what prior methods for unsupervised KG-QA
use~\cite{diefenbach2019qanswer}.
For incomplete and ungrammatical follow-up questions, we capture
context in the form of a subgraph as well, and we dynamically maintain
it as the conversation proceeds.
This way, relevant entities and predicates from previous turns are kept in 
the gradually expanding context.
However, we need to be careful about growing the subgraph too much as the
conversation branches and broadens in scope. As nodes in a KG have
many 1-hop neighbors and a huge number of 2-hop neighbors,
there is a high risk of combinatorial explosion, and a huge subgraph would
no longer
focus on the topic of interest.
\convex copes with this critical issue by judiciously expanding
the context subgraph,
using a combination of look-ahead, weighting, and pruning techniques.
Hence the ``look before you hop'' in the paper title.

Specifically, \convex works as follows. Answers to the first question are 
obtained
by any standard KG-QA system (we use the state-of-the-art
system QAnswer~\cite{diefenbach2019qanswer} and other variants in our 
experiments over Wikidata).
Entities in the initial question $q^0$, the answer $a^0$, and their connections 
initialize
a context subgraph $X^1$ ($X^t$ for turn $t$) for the conversation in the KG. 
When a follow-up
question $q^1$ arrives, all nodes (entities, predicates, or types) in the 
KG-neighborhood of $X^1$ are deemed as
candidates that will be used to expand the current graph.
Brute force
addition of all neighbors to $X^t$ will quickly lead to an explosion
in its size after a few turns
(hugely exacerbated if popular entities are added, e.g. \struct{Germany} and
\struct{Barcelona} have
$\simeq 1.6M$ and $\simeq40k$ neighbor entities in Wikidata).
Thus, we opt for prudent expansion as follows. Each neighbor is 
scored based on
its similarity to
the question,
its distance to important nodes
in $X^1$, the conversation turn $t$, and KG priors. This information is stored
in
in respective sorted lists with these neighbors as elements.

A small number of top-scoring neighbors of $X^1$ in a turn, termed 
``frontier nodes'' ($\{F^1\}$),
are identified by aggregating information across these queues. Next, all KG 
triples (SPO facts) for these frontiers \textit{only}, are added to $X^1$,
to produce an expanded context $X^1_+$. 
These
$\{F^1\}$ are the most relevant nodes w.r.t the current question $q^1$, and
hence are expected to contain the answer $a^1$ in their close proximity. Each 
entity
in $X^1_+$ is thus scored by its distance to each frontier node $F^1_i$ 
and other
important nodes in $X^{1}$, and the top-ranked entity
(possibly multiple, in case of ties) is returned as $a^1$.
This process is then iterated for each turn in the conversation with 
question $q^t$, producing
$X^t$, $X^t_+$, $\{F^t\}$, and ultimately $a^t$ at each step.

\textbf{Benchmark.} We 
compiled the first realistic benchmark, termed \convquestions,
for conversational KG-QA.
It contains about $11k$ 
conversations which can be evaluated over Wikidata.
They are compiled from the inputs of crowdworkers on
Amazon Mechanical Turk,
with
conversations from five domains: \phrase{Books}, \phrase{Movies},
\phrase{Soccer}, \phrase{Music}, and \phrase{TV Series}.
The questions feature a variety of complex question phenomena like comparisons,
aggregations, compositionality, and
temporal reasoning. 
Answers are grounded in 
Wikidata entities to enable fair comparison across diverse methods.

\textbf{Contributions.} 
The main contributions of this work are:
\squishlist
	\item We devise \convex, 
an
unsupervised method for addressing 
	conversational question answering over knowledge graphs.
	\item We release \convquestions, the first realistic benchmark to evaluate 
	conversational KG-QA.
	\item We present extensive experiments, showing how \convex
	enables any stand-alone system with conversational support.
	\item An online demo and all code, data and results is available at
\textbf{\texttt{http://qa.mpi-inf.mpg.de/convex/}}.
\squishend


\section{Concepts and Notation}
\label{sec:concepts}

\begin{figure*} [t]
	\centering
	\includegraphics[width=\textwidth]{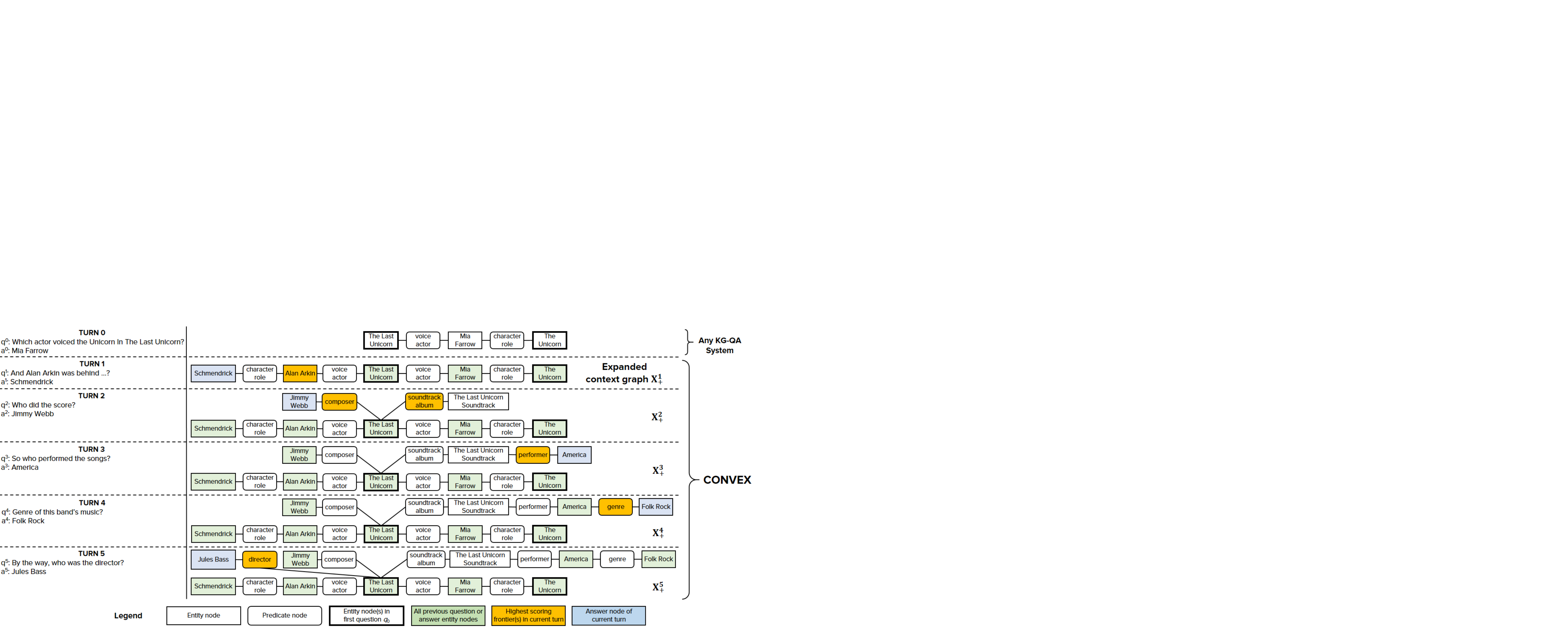}
	\caption{A typical conversation illustrating perfect (but simplified) 
		context expansion and answering at every turn.}
	\label{fig:convex}
	\vspace*{-0.4cm}
\end{figure*}


We first introduce concepts that will assist in an easier
explanation for the \convex method, and corresponding notations. An 
example workflow instantiating these concepts is shown in Fig.~\ref{fig:convex},
and Table~\ref{tab:notation} provides a ready reference.

\begin{table} [t]
	\centering
	\resizebox*{\columnwidth}{!}{
		\begin{tabular}{l l}
			\toprule
			\textbf{Notation}				& \textbf{Concept}										\\ \toprule
			$K, E, P, \mathcal{C}, L$		& Knowledge graph, entity, predicate, class, literal	\\
			$S, P, O$						& Subject, predicate, object							\\
			$N, \mathcal{E}$				& Nodes and edges in graph								\\ \midrule
			$C, t$							& Conversation, turn									\\
			$q^t, a^t$						& Question and answer at turn $t$						\\
			$X^t, X^t_+$					& Initial and expanded context graphs at turn $t$ 		\\
			$\mathcal{N}(X^t)$				& $k$-hop neighborhood of nodes in $X^t$				\\
			$F^t = \{F^t_1 \ldots F^t_r\}$	& Frontier nodes at turn $t$							\\ 
			$E(q^t)$						& Entities mapped to by words in $q^t$ 					\\ \bottomrule
	\end{tabular}}
	\caption{Notation for key concepts in \convex.}
	\label{tab:notation}
	\vspace*{-0.9cm}
\end{table}

\textbf{Knowledge graph.} A knowledge graph, or a knowledge base, $K$ is a set 
of subject-predicate-object
(SPO) RDF triples, each representing a real-world fact, where $S$ is of 
type entity $E$
(like \struct{The Last Unicorn}), $P$ is a predicate (like \struct{director}), 
and
$O$ is another entity, a class $\mathcal{C}$ (like
\struct{animated feature film}), 
or a literal $L$
(like \struct{19 November 1982}). All $E$, $P$, $\mathcal{C}$, and $L$ in $K$ 
are canonicalized. Most modern KGs support 
$n$-ary facts like movie-cast information
(with more than two $E$ and more than one $P$)
via reification with intermediate nodes~\cite{suchanek2007yago}. In 
Wikidata, such information is
represented via optional qualifiers with the main fact (\struct{TheLastUnicorn 
	voiceActor MiaFarrow . characterRole TheUnicorn}). Compound Value Types
(CVTs) were the Freebase analogue. Tapping into 
qualifier information is a challenge for SPARQL queries, but is easily 
accessible in a graph-based method like \convex.

\convex stores the KG as a \textit{graph} $K = (N, \mathcal{E})$, with a 
set of nodes 
$N$ and a set of edges $\mathcal{E}$, instead of
a database-like RDF triple store. Each $E$, $P$, $\mathcal{C}$, and $L$ is 
assigned 
a unique node in $K$, with two nodes $n_1, n_2 \in N$ having an edge
$e \in \mathcal{E}$ between 
them if there is a triple $\langle n_1, n_2, \cdot \rangle\in K$ or
$\langle \cdot, n_1, n_2 \rangle \in K$. While it is more standard practice to
treat each $P$ as an edge label, we represent every item in $K$ as a node,
because it facilitates computing standard graph measures downstream.
Examples of sample $n$ and
$e$ are shown in the (sub-)graphs in Fig.~\ref{fig:convex}. $E$ and $P$ nodes 
are in rectangles with sharp and rounded corners, respectively. For simplicity, 
$\mathcal{C}$ and $L$ nodes are not shown. An important thing to note is that
each instance of some $P$ retains an \emph{individual existence} in the graph 
to 
prevent false inferences (e.g. two \struct{voice actor} nodes in the figure).
As a simple example, if we merge the node for \struct{married}
from two triples $\langle E_1, married, E_2 \rangle$ and
$\langle E_3, married, E_4 \rangle$, then we may accidentally infer that $E_1$ 
is married to $E_4$ during answering.

\textbf{Conversation.} A conversation $C$ with $T$ turns is made up of a 
sequence of questions
$Q = \{q^t\}$ and corresponding answers $A = \{a^t\}$, where $t = 0, 1, \ldots 
T$, such that $C = \langle (q^0, a^0), (q^1, a^1) \ldots (q^T, a^T)\rangle$.
Fig.~\ref{fig:convex} (left side) shows a typical $C$ that \convex handles,
with $T = 5$ (six turns). Usually, $q^0$ is well-formed, and all other $q^t$
are ad hoc.

\textbf{Question.} Each $q^t$ is a sequence
of words $q^t_i$, such that $q_t = \langle {q^t_1} \ldots {q^t_{|q^t|}}\rangle$, 
where
$|q^t|$ is the number of words in $q^t$. During answering, each word in 
$q^t$ potentially maps
to one or more items in $K$ ($q^t_i \mapsto E \cup P \cup \mathcal{C} \cup L$). 
However, since
conversations revolve around entities of interest, we fixate on the mapped 
entities, and refer to them as $E(q^t)$. E.g.,
\phrase{Alan} in $q^1 \mapsto \{\struct{Alan Arkin}\}$, and 
\phrase{score} in
$q^2 \mapsto \{\struct{composer, soundtrack, The Last Unicorn Soundtrack}\}$;
so $E(q^1) = \{\struct{Alan Arkin}\}$, and
$E(q^2) = \{\struct{The Last Unicorn Soundtrack}\}$.

\textbf{Answer.} Each answer $a^t$ to question $q^t$ is
a (possibly multiple, single, or null-valued) \textit{set} of entities or
literals in $K$, i.e. $a^t \in E \cup L$ 
(questions asking for predicates or 
classes are usually not realistic). Each $a^t$ is shaded in light blue in
Fig.~\ref{fig:convex}.

\textbf{Context subgraph.} In the \convex model, every turn $t$ in $C$ is
associated with a context $X^t$, that is a subgraph grounded or anchored in a 
localized zone in $K$. Each $X^t$ subgraph consists of:
(i) the previous question entities in $C$, $\{E(q^1) \cup \ldots E(q^{t-1})\}$,
(ii) previous answer entities in $C$: $\{a^1 \cup \ldots a^{t-1}\}$,
(iii) intermediate nodes and edges connecting the above in $K$. All $E$ nodes
corresponding to turns $1, \ldots, (t-1)$, are shaded in light green.

\textbf{Frontier nodes.} At every turn $t$, nodes in the $k$-hop neighborhood
of $X^t$, $\mathcal{N}(X^t)$,
define something like a \textit{border} to which we may need to \textit{expand} 
$X^t$ 
for answering the next $q^t$ (current nodes in 
$X^t$ are subsumed in $\mathcal{N}(X^t)$). The number of hops $k$ is 
small in practice, owing to the 
fact that typical users do not suddenly make large topic jumps during a 
specific $C$.
Even then, since expanding $X^t$ to include every $n \in 
\mathcal{N}(X^t)$ 
results in
an exponential growth rate for its size that we wish to avoid, we first select
the best (top-$r$) nodes in $\mathcal{N}(X^t)$.
These optimal 
expansion points in $\mathcal{N}(X^t)$ are referred to as
\emph{frontier nodes}, a ranked set $F^t = \{F^t_1, \ldots F^t_r\}$, and are 
the most 
relevant nodes with respect to the current question $q^t$ and
the current context $X^t$,
as ranked by some \textit{frontier score} (defined later).
This entails that only those triples (along with qualifiers)
(analogously, the resultant nodes and edges) that connect these $F^t_i$ 
to the $X^t$ are added to the context.
The top-$1$
frontier node $F^t_1$ at every $t$ is shown in orange in the figure (multiple 
in case of ties).

\textbf{Expanded context.} Once all triples in $K$ corresponding to frontier
nodes $F^t$ are added to $X^t$, we obtain an expanded context graph $X^t_+$.
All nodes in $X^t_+$ are candidate answers $a_t$, that are scored
appropriately.
Fig.~\ref{fig:convex} shows expanded contexts $\{X^t_+\}$ for every $t$ in
our example conversation. Corresponding $X^t$ can be visualized by removing
facts with the orange frontiers. Notably, $X^t_+ = X^{t+1}$.

\section{The \convex algorithm}
\label{sec:method}

We now describe the \convex conversation handler method, that can be envisaged 
as a seamless plug-in enabling stand-alone KG-QA systems to answer
incomplete follow up questions with possibly ungrammatical and informal 
formulations. \convex thus requires an underlying KG, a standard QA system that 
can answer well-formulated questions, and the conversational utterances as
input. On receiving an input question at a given turn, our method proceeds in
two stages: (i) expand the context graph, and (ii) rank the answer candidates
in the expanded graph. We discuss these steps next.

\vspace*{-0.1cm}
\subsection{Context expansion}
\label{subsec:con-exp}

The initial question $q^0$ is answered by the KG-QA system that \convex
augments, and say, that it produces answer(s) $a^0$. Since entities in the
original question are of prime importance in a conversation, we use
any off-the-shelf named entity recognition and disambiguation (NERD) system
like TagMe~\cite{ferragina2010tagme} or
AIDA~\cite{hoffart2011robust} to identify entities $E(q^0)$.
Such $E(q^0)$, $a^0$, and
the KG connections between them initialize the context subgraph $X^1$.

Now, when the first question $q^1$ arrives, we need to look for answer(s) $a^1$
in the vicinity of $X^1$. The main premise of this work is not to treat 
\textit{every node}
in such neighborhood of $X^1$, and more generally, $X^t$, as an answer
candidate. This is because, over turns, expanding the context, by any means, is
inevitable: users can freely drift away and revisit the initial entities
of interest over the full course of the conversation. Under this postulate,
the total number of 
such context nodes can easily go to the order of \textit{millions}, aggravated 
by the 
presence of popular entities, especially countries (\struct{UK, Russia}) or 
cities (\struct{Munich, Barcelona}) in the KG around prominent entities of 
discussion (\struct{Harry Potter, Christiano Ronaldo}).

The logical course of action, then, is to perform this expansion in a somewhat
austere fashion, which we propose to do as follows. We wish to identify some
key nodes in the $k$-hop neighborhood of $X^1$, that will prove the most 
worthwhile if included into $X^1$ (along with their connections to $X^1$)
w.r.t. answering $q^1$. We call these optimal expansion points
\textit{frontier nodes}. From here on, we outline frontier identification at 
a general conversation turn $t$, where $t = 0, 1, \ldots T$. Frontiers are
marked by three signals: (i) relevance to the words in $q^t$; (ii) relevance
to the current context $X^t$; and (iii) KG priors. We now explain these 
individual factors.

\begin{figure} [t]
	\centering
	\includegraphics[width=0.9\columnwidth]{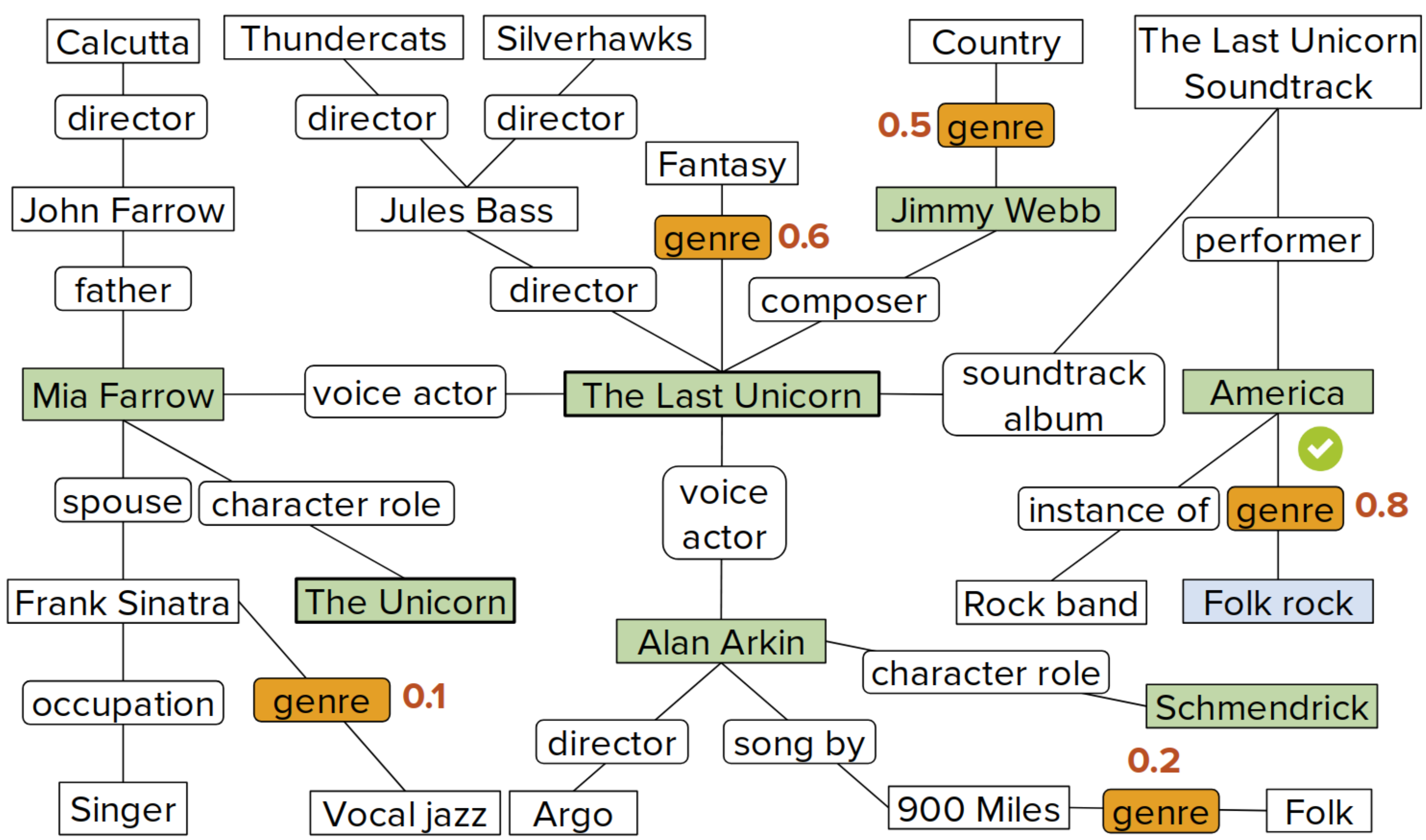}
	\caption{An illustration of the ambiguity in frontier node selection
		for a specific question word (\phrase{genre} in
		$q^4 = $ \utterance{Genre of this band's music?}), and how effective 
		scoring 
		can potentially pick the best candidate in a noisy context graph $X^4$.}
	\label{fig:genre}
	\vspace*{-0.3cm}
\end{figure}

\textbf{Relevance to question.} The question words $\{q^t_i\}$ provide a direct
clue to the relevant nodes in the neighborhood. 
However, there is often a vocabulary mismatch between what users specify in their
questions and the KG terminology, as typical users are unaware of the KG schema.
For example, let us consider $q^3 =$ \utterance{Who did the score?}.
This indicates the sought information is about the score of the movie but
unfortunately the KG does not use this term. So, we define
the \textit{matching similarity score} of
a neighbor $n$ with a question word using cosine similarity between
word2vec~\cite{mikolov2013distributed}
embeddings of the 
node label and the word. \textit{Stopwords} like
\textit{and, of, to, etc.} are excluded from 
this similarity.
For multiword phrases, we use
an averaging of the
word vectors~\cite{wieting2016towards}. The cosine similarity is originally
in $[-1, +1]$: it is scaled to $[0, 1]$ using min-max normalization
for comparability to the later measures. So we have:
\begin{equation}
	match(n, q^t_i | n \in \mathcal{N}(X^t))
	= cos_{norm}(w2v(label(n)), w2v(q^t_i))
	\label{eq:matchw}
\end{equation}
We then take the maximum of these word-wise scores to define the matching
score of a candidate frontier to the question as a whole:
\begin{equation}
	match(n, q^t | n \in \mathcal{N}(X^t))
	= \max_i match(n, q^t_i)
	\label{eq:match}
\end{equation}
\textbf{Relevance to context.} Nevertheless, such soft lexical matching
with embeddings is hardly
enough. Let us now consider the word \phrase{genre} in
$q^3 =$ \utterance{Genre of this band's music?}. Looking at the toy example in
Fig.~\ref{fig:genre},
we see that even with an exact match, there are five \struct{genre}-s lurking in
the vicinity at $X^4$ (there are several more in reality), where the one
connected to \struct{America} is the intended fit.

We thus define the relevance
of a node $n$ to the current context $X^t$ as the total graph distance $d_K$
(in number of hops in $K$) of $n$ to the nodes in $X^t$. Note that
we are interested
in the relevance score being directly proportional to this quantity, and hence
consider proximity, the reciprocal of distance ($1/d_K$),
as the measure instead.
For the aggregation over nodes $x$ in $X^t$, we prefer
$\sum (1/d_K(n, x))$
over $1/\sum(d_K(n, x))$ as the latter is more sensitive to outliers.

Next, not all nodes in $X^t$ are valuable for the answering process.
For anchoring a conversation $C$ in a KG, 
entities that have appeared in \textit{questions} or as \textit{answers}
in turns
$0, \ldots, (t-1)$, are what specifically matter. Thus, it suffices to consider
only $\bigcup_{j = 0}^{t-1} E(q^j)$ and $\bigcup_{j = 0}^{t-1} a^j$ for
computing the
above proximities. We encode this factor using an indicator function
$\mathbb{1}_{QA}(x | x \in X^t)$ that equals $1$ if
$x\in \bigcup_{j = 0}^{t-1} [E(q^j) \cup a^j]$, 
and zero otherwise.

Contributions of such Q/A nodes
$x \in X^t$ ($E(q^j)$ or $a^j$) should be weighted according to the turn
in which they appeared in their respective roles, denoted by $turn(x)$).
This is
when such nodes had the ``spotlight'', in a sense; so recent turns have
higher weights
than older ones. In addition, since the \emph{entity in the
first question} $E(q^0)$ may always be important as the theme of the conversation
(\struct{The Last Unicorn}),
$turn(E(q^0))$ is set to the maximum value $(t-1)$ instead of zero.
We thus define the \textit{context proximity
score} for neighbor $n$, normalized by the number of Q/A nodes in
the context, as:
\begin{equation}
	prox(n, X^t | n \in \mathcal{N}(X^t))
	= \frac{\sum\limits_{x \in X^t}turn(x)
	\cdot \mathbb{1}_{QA}(x) \cdot 1/d_K(n, x)}
	{\sum\limits_{x \in X^t} \mathbb{1}_{QA}(x)}
	\label{eq:prox}
\end{equation}
\textbf{KG priors.} Finally, KG nodes have inherent salience (or prominence)
values that reflect their likelihoods of being queried about in users' questions.
For example, \struct{Harry Potter} has higher salience as opposed to some
obscure book like \struct{Harry Black}, and the same can be said to hold
about the \struct{author} predicate
compared to \struct{has edition} for books. Ideally, such salience should be
quantified using large-scale query logs from real users that commercial Web
search engines possess. In absence of such resources, we use a more intrisic
proxy for
salience: the frequency of the concept in the KG. The raw frequency
is normalized
by corresponding maximum values for entities, predicates, classes, and literals,
to give $freq_{norm}$. Thus, we have the \textit{KG prior} for a node $n$ as:
\begin{equation}
	prior(n, K) = freq_{norm}(n, K)
	\label{eq:prior}
\end{equation}
\textbf{Aggregation using Fagin's Threshold Algorithm.}
We now have three independent signals from the question, context, and the KG
regarding the likelihood of a node being a frontier at a given turn.
We use the \textit{Fagin's Threshold Algorithm (FTA)}~\cite{fagin2003optimal}
to aggregate items in the three sorted lists
$\mathcal{L} = \{\mathcal{L}_{match}, \mathcal{L}_{prox}, \mathcal{L}_{prior}$\},
that are created when candidates are scored by each of these signals.
FTA is chosen as it is an optimal algorithm with correctness and performance
guarantees for rank aggregation.
In FTA, we perform sorted access in parallel to each of the three
sorted lists $\mathcal{L}_i$. As each
candidate frontier node $n$ is seen under sorted access, we retrieve each of its
individual scores by random access. We then compute a frontier score for $n$
as a simple linear combination of
the outputs from Eqs.~\ref{eq:match}, \ref{eq:prox},
and \ref{eq:prior} using
frontier hyperparameters $h_1^F$, $h_2^F$, and $h_3^F$, 
where $\sum_i h_i^F = 1$:
\vspace*{-0.5cm}
\begin{multline}
frontier(n, q^t | n \in \mathcal{N}(X^t)) \\
= h_1^F \cdot match(n, q^t)
+ h_2^F \cdot prox(n, X^t)
+ h_3^F \cdot prior(n, K)
\label{eq:frontier}
\end{multline}
In general, FTA requires a monotonic score aggregation function
$frontier()$
such that
$frontier(s_1, s_2, s_3) \leq frontier(s'_1, s'_2, s'_3)$
whenever $s_i \leq s'_i, \; \forall \; i$, where
the component scores of $frontier(n)$ are denoted as $s_i$'s
(Eq.~\ref{eq:frontier}) in corresponding lists $\mathcal{L}_i$.
Once the above is done, as nodes are accessed from $\mathcal{L}_i$, if
this is one of the top $r$ answers so far, we remember it. 
Here, we assume a buffer of bounded size.
For each $\mathcal{L}_i$, let
$\hat{s_i}$ be the score of the last node seen under sorted
access. We define the threshold value $thresh$ to be
$frontier(\hat{s_1}, \hat{s_2}, \hat{s_3})$.
When
$r$ nodes have been seen whose frontier score is at least
$thresh$, then we stop and return the top $r$ nodes as the final frontiers.


Thus, at the end of this step,
we have a set of $r$ frontier nodes $\{F^t_i\}_{i = 1}^r$ for turn $t$. If any
of these frontiers are entities, they are used to populate $E(q^t)$.
We add the
triples (along with qualifiers if any) that connect the $F^t_i$
to the current context $X^t$ to produce the expanded graph $X^t_+$
(the step containing $facts(F^t)$ in Algorithm~\ref{algo:convex}).

\begin{algorithm} [t]
	\SetAlgoLined
	$t = 1$\;
	\textbf{initialize} $X^1 = E(q^0) \cup a^0 \cup K(E(q^0), a^0)$\;
	\While{$t \leq T$}
	{		
		\For{$n \in \mathcal{N}(X^t)$}
		{
			\textbf{compute} $match(n, q^t)$ [Eq.~\ref{eq:match}]\;
			\textbf{compute} $prox(n, X^t)$ [Eq.~\ref{eq:prox}]\;
			\textbf{compute} $prior(n, K)$ [Eq.~\ref{eq:prior}]\;
			\textbf{insert} scores into sorted lists $\mathcal{L} = \{\mathcal{L}_{match}, \mathcal{L}_{prox}, \mathcal{L}_{prior}\}$\;
		}
		\textbf{find} $\{F^t_i\}^{i = 1}_r$  = Fagin's-Threshold-Algorithm($\mathcal{L}, r$) [Eq.~\ref{eq:frontier}]\;
		\textbf{assign} $E(q^t) = E(F^t)$\;
		\textbf{expand} $X^t_+ = X^t \cup facts(F^t)$\;
		
		\For{a $\in X^t_+$}
		{
			\textbf{compute} $ans\!-\!score(a$) [Eq.~\ref{eq:ans}]\;
		}
		\textbf{find} $a^t = \arg \max_{X^t_+} ans\!-\!score(a)$\;
		$X^{t+1} \leftarrow X^t_+$\;
		$t \leftarrow t + 1$\;
	}
	\KwResult{$\langle a^1, \ldots a^T \rangle$}	
	\caption{\convex($K, T, q^0, a^0, E(q^0), \langle q^1, \ldots q^T \rangle, r$)}
	\label{algo:convex}
\end{algorithm}
\setlength{\textfloatsep}{0.1cm}
\setlength{\floatsep}{0.1cm}

\vspace*{-0.2cm}
\subsection{Answer ranking}
\label{subsec:ans-rank}

Our task now is to look for answers $a_t$ in $X^t_+$.
Since frontiers are the most relevant nodes in $X^t_+$ w.r.t question $q_t$,
it is expected that $a_t$ will appear in their close proximity.

However, labels of frontier nodes only reflect what was \textit{explicit} in
$q^t$, the unspecified or \textit{implicit} part of the context
in $q^t$ usually refers to a previous question or answer entity
($\bigcup_{j = 0}^{t-1} [E(q^j) \cup a^j]$). Thus, we should consider closeness
to these context entities as well. Note that just as before, frontiers and
Q/A entities both come with corresponding weights: frontier scores and turn id's,
respectively. Thus, while considering proximities is key here, using weighted
versions is a more informed choice. We thus score every node $a \in X^t_+$ by
its weighted proximity, using Eqs.~\ref{eq:prox} and \ref{eq:frontier},
as follows (again, we invert distance to use a measure \textit{directly
proportional} to the candidacy of an answer node):
\begin{multline}
	a^t = \arg \max_{a \in X^t_+} [
	h_1^A \cdot \frac{\sum_{i = 1}^r [frontier(F^t_i)
		\cdot (1 / d_K(a, F^t_i))]}{r} 
\\	+ h_2^A \cdot \frac {\sum_{x \in X^t} turn(x) \cdot \mathbb{1}_{QA}(x)
	\cdot 1/d_K(a, x)}{\sum\limits_{x \in X^t} \mathbb{1}_{QA}(x)}] 
	\label{eq:ans}
\end{multline}
Contributions by proximities to frontier and Q/A nodes
(each normalized appropriately) are again combined
linearly with answer hyperparameters $h^A_1$ and $h^A_2$, where 
$h^A_1 + h^A_2 = 1$. Thus, the final answer score also lies in $[0, 1]$.
Finally, the top scoring $a^t$ (possibly multiple, in case of ties)
node(s) is returned as the \textit{answer} to $q^t$.

The \convex method
is outlined in Algorithm~\ref{algo:convex}. As mentioned before, $a^0$ and
$E(q^0)$ are obtained by passing $q^0$ through a stand-alone KG-QA system,
and a NERD algorithm, respectively. $K(\cdot)$ returns all KG triples that
contain the arguments of this function, and the generalized $E(\cdot)$ returns
the set of entities from its arguments. Note that this algorithm illustrates
the workings of \convex in a static setting when all $q^t (t = 0 \ldots T)$
are given upfront;
in a real setting, each $q^t$ is issued interactively with a user in the loop.

\section{The \textsc{ConvQuestions} Benchmark}
\label{sec:benchmark}

\begin{table} [t] \small
	\centering
	\begin{tabular}{p{2.5cm} p{5cm}}
		\toprule
		\textbf{Attribute}		& \textbf{Value}												\\ \toprule
		Title					& Generate question-answer conversations on popular entities	\\
		Description				& Generate conversations in different domains (books, movies,
		music, soccer, and TV series) on popular entities of your choice. You need to
		ask natural questions and provide the corresponding answers via Web search.			\\ \midrule
		Total participants		& $70$ 														\\
		Time allotted per HIT	& $6$ hours 												\\
		Time taken per HIT 		& $3$ hours 												\\
		Payment per HIT			& $25$ Euros 												\\ \bottomrule
	\end{tabular}
	\caption{Basic details of the AMT HIT (five conversations).}
	\label{tab:hit}
	\vspace*{-0.3cm}
\end{table}

\begin{table*} [t] \small
	\centering
	\resizebox*{\textwidth}{!}{
		\begin{tabular}{p{0.5cm} | p{3cm} | p{3cm} | p{3cm} | p{3cm} | p{3cm}}
			\toprule
			\textbf{Turn}	& \textbf{Books}	&	\textbf{Movies}	&	\textbf{Soccer}	&	\textbf{Music}	&	\textbf{TV series} \\ \toprule
			$q^0$			& \utterance{When was the first book of the book series The Dwarves published?}	&	\utterance{Who played the joker in The Dark Knight?}	&	\utterance{Which European team did Diego Costa represent in the year 2018?}	&	\utterance{Led Zeppelin had how many band members?}	&	\utterance{Who is the actor of James Gordon in Gotham?} \\
			$a^0$			& \struct{2003}	&	\struct{Heath Ledger}	&	\struct{Atl\'{e}tico Madrid}	&	\struct{4}	&	\struct{Ben McKenzie} \\ \midrule
			$q^1$			& \utterance{What is the name of the second book?}	&	\utterance{When did he die?}	&	\utterance{Did they win the Super Cup the previous year?}	&	\utterance{Which was released first: Houses of the Holy or Physical Graffiti?}	&	\utterance{What about Bullock?} \\
			$a^1$			& \struct{The War of the Dwarves}	&	\struct{22 January 2008}	&	\struct{No}	&	\struct{Houses of the Holy}	&	\struct{Donal Logue} \\ \midrule
			$q^2$			& \utterance{Who is the author?}	&	\utterance{Batman actor?}	&	\utterance{Which club was the winner?}	&	\utterance{Is the rain song and immigrant song there?}	&	\utterance{Creator?} \\
			$a^2$			& \struct{Markus Heitz}	&	\struct{Christian Bale}	&	\struct{Real Madrid C.F.}	&	\struct{No}	&	\struct{Bruno Heller} \\ \midrule
			$q^3$			& \utterance{In which city was he born?}	&	\utterance{Director?}	&	\utterance{Which English club did Costa play for before returning to Atl\'{e}tico Madrid?}	&	\utterance{Who wrote those songs?}	&	\utterance{Married to in 2017?} \\
			$a^3$			& \struct{Homburg}	&	\struct{Christopher Nolan}	&	\struct{Chelsea F.C.}	&	\struct{Jimmy Page}	&	\struct{Miranda Cowley} \\ \midrule
			$q^4$			& \utterance{When was he born?}	&	\utterance{Sequel name?}	&	\utterance{Which stadium is this club's home ground?}	&	\utterance{Name of his previous band?}	&	\utterance{Wedding date first wife?} \\ 
			$a^4$			& \struct{10 October 1971}	&	\struct{The Dark Knight Rises}	&	\struct{Stamford Bridge}	&	\struct{The Yardbirds}	&	\struct{19 June 1993} \\ \bottomrule
	\end{tabular}}
	\caption{Representative conversations in \convquestions from each domain, highlighting the stiff challenges they pose.}
	\label{tab:convq}
	\vspace*{-0.7cm}
\end{table*}

\subsection{Benchmark creation}
\label{subsec:create}

\textbf{Limitations of current choices.} Popular benchmarks for
KG-QA like WebQuestions~\cite{berant2013semantic},
SimpleQuestions~\cite{bordes2015large},
WikiMovies~\cite{miller2016key},
ComplexWebQuestions~\cite{talmor2018web},
and
ComQA~\cite{abujabal19comqa},
are all designed for one-shot answering
with well-formulated
questions. The CSQA dataset~\cite{saha2018complex} takes preliminary steps
towards the sequential KG-QA
paradigm, but it is extremely artificial: initial and follow-up questions
are generated semi-automatically via templates, and sequential utterances
are only simulated by stitching questions with shared entities or relations
in a thread, without a logical flow. 
QBLink~\cite{elgohary2018dataset}, 
CoQA~\cite{reddy2018coqa}, 
ans ShARC~\cite{saeidi2018interpretation} are recent resources for sequential QA
\textit{over text}. The SQA resource~\cite{iyyer2017search}, derived
from WikiTableQuestions~\cite{liang2015compositional}, is aimed at driving
conversational
QA over (relatively small) Web \textit{tables}.

\textbf{Conceptual challenges.} In light of such limitations, we overcome
several conceptual challenges to build
the first realistic benchmark for conversational KG-QA, anchored in Wikidata.
The key questions included, among others: Should we choose $q^0$ from existing
benchmarks and ask humans to create only follow-ups? Should the answers already
come from some KG-QA system, observing which, users create follow-ups?
Should we allocate templates
to crowdworkers to systematically generate questions that miss either entities,
predicates, and types? Can we interleave questions by different workers to create
a large number of conversations? Can we permute the order of follow-ups to
generate an even larger volume? If there are multiple correct $a^t$, and
$q^{t+1}$ in the benchmark involves a different $a^t$ than what the system
returns at run-time, how can we evaluate such a dynamic workflow? How can we built a KG-QA
resource that is faithful to the setup but is not overly limited to the
information the KG contains today?

\textbf{Creating \convquestions.} With insights from a meticulous in-house pilot
study with ten students over two weeks, we posed the conversation generation task
on Amazon Mechanical Turk (AMT)
in the most
\textit{natural setup}:
Each crowdworker was asked to build a conversation by asking five 
sequential questions \textit{starting} from any \textit{seed} entity of his/her
choice, as this is an intuitive mental model that humans may have when satisfying
their real information needs via their search assistants.
A system-in-the-loop is hardly ideal: this creates comparison across methods
challenging, is limited by the shortcomings of the chosen system, and most
crucially, there exist no such systems today with satisfactory performance.
In a single AMT Human Intelligence Task (HIT), Turkers had to create one
conversation each from five domains: \phrase{Books}, \phrase{Movies},
\phrase{Soccer}, \phrase{Music}, and \phrase{TV Series} (other potential choices
were \phrase{Politics}, but we found that it quickly becomes subjective,
and \phrase{Finance}, but that
is best handled by relational databases and not curated KGs). Each conversation
was
to have five turns, including $q^0$. To keep conversations as \textit{natural}
as possible, we neither interleaved questions from multiple Turkers, nor permuted
orders of questions within a conversation. For quality, only AMT Master Workers
(who have consistently high performances:
see \url{https://www.mturk.com/help#what_are_masters}),
were allowed to participate.
We registered $70$ participants,
and this resulted in $350$
initial conversations, $70$ from each domain.

Along with questions,
Turkers were asked to provide textual surface forms and Wikidata links of
the seed entities and
the \textit{answers} (via Web search), along with \textit{paraphrases} of each
question.
The paraphrases provided us with two versions of the same question,
and hence a means of augmenting the core data with
several interesting variations that can simultaneuosly boost and test
the robustness of KG-QA systems~\cite{dong2017learning}. Since paraphrases of
questions (any $q^t$) are
always semantically equivalent and interchangeable, each conversation with
five turns thus resulted in $2^5 = 32$ distinct conversations (note that this
\textit{does not} entail shuffling sequences of the utterances).
Thereby, in total,
we obtained $350 \times 32 = 11,200$ such conversations, that we release
with this paper.

If the answers were dates or literals like measurable quantities with units,
Turkers were asked to follow the Wikidata formats for the same.
They were 
provided with \textit{minimal syntactic guidelines} to remain natural in
their questions.
They were shown judiciously selected examples so as not to ask
opinionated questions (like \utterance{best film by this actor?}), or
other non-factoid questions (causal, procedural, etc.).
The authors invested substantial manual effort for quality control 
and spam prevention, by verifying both answers of random utterances,
and alignments between provided texts and Wikidata URLs.
Each question was allowed to have at most three answers, but single-answer
questions were encouraged to preclude the possibility of 
non-deterministic workflows during evaluation.

To allow for \convquestions being relevant for a few years into the future,
we encouraged users to ask complex questions involving joins, comparisons,
aggregations, temporal information needs, and so on. Given the complexity
arising from incomplete cues, these additional facets pose an even greater
challenge for future KG-QA systems. 
So as not to restrict questions to only those predicates that are present in 
Wikidata 
today, relations connecting question and answer entities are sometimes missing
in the KG but can be 
located in sources like Wikipedia, allowing scope for both future growth of
the KG, and experimentation with text plus KG combinations.

\subsection{Benchmark analysis}
\label{subsec:analysis}

Basic details of our AMT HIT are provided in Table~\ref{tab:hit} for reference.
Question entities and expected answers had a balanced distribution among
human (actors, authors, artists) and non-human types (books, movies, stadiums).
Detailed distributions are omitted due to lack of space.
Illustrative examples of challenging questions from \convquestions are in
Table~\ref{tab:convq}. We see manifestations of: incomplete cues (TV Series;
$q^3$), ordinal questions (Books; $q^1$), comparatives
(Music; $q^1$), indirections (Soccer; $q^4$), anaphora (Music; $q^3$),
existentials (Soccer; $q^2$), temporal reasoning (Soccer; $q^3$), among other
challenges.
The average lengths of the first and follow-up questions were $9.07$ and 
$6.20$ words, respectively. Finally, we present the key quantifier for
the difficulty in our benchmark: the average KG distance of answers from
the original seed entity is $2.30 \; hops$, while the highest goes up to as high
as \textit{five} KG hops. Thus, an approach that remains fixated on
a specific entity is doomed to fail:
context expansion is the key to success on \convquestions.

\section{Experimental Setup}
\label{sec:setup}

\subsection{Baselines and Metrics}
\label{subsec:baselines}

\textbf{Stand-alone systems.} We use the state-of-the-art system
QAnswer~\cite{diefenbach2019qanswer}, and also Platypus~\cite{tanon2018demoing},
as our stand-alone KG-QA systems, that serve as baselines, and which we enhance
with \convex. At the time of writing (May 2019), these are the only two systems
that have running prototypes over Wikidata.

To make \convex a self-sufficient
system, we also implement a \textit{na\"{i}ve} version of answering
the first question as follows.
Entities are detected in $q^0$ using the TagMe NERD
system~\cite{ferragina2010tagme}, and mapped to their Wikidata IDs via Wikipedia
links provided by TagMe. Embeddings were obtained by averaging word2vec vectors
of the non-entity words in $q^0$, and their cosine similarities were
computed around
each of the predicates around the detected entities $E(q^0)$.
Finally, the best $(E, P)$
pair was found (as a joint disambiguation), and the returned answer was
the subject
or object in the triple according as the triple structure was
$\langle \cdot, P, E\rangle$ or $\langle E, P, \cdot \rangle$.

Due to the complexity in even the first question $q^0$ in \convquestions,
all of the above systems achieve a very poor performance for $a^0$ on
the benchmark.
This limits the value that \convex can help these systems achieve, as $E(q^0)$
and $a^0$ together initialize $X^1$. To decouple the effect of the original QA
system, we experiment with an \textit{Oracle} strategy, where we use $E(q^0)$
and $a^0$ provided by the human annotator (Turker who created the conversation).

\textbf{Conversation models.} As intuitive alternative strategies to \convex for
handling conversations, we explore two variants: (i) the \textit{star-join}, and
(ii) the \textit{chain-join} models. The naming is inspired by DB terminology,
where
a star query looks for a join on several attributes around a single variable
(of the form \struct{SELECT ?x WHERE \{?x att$_1$ val$_1$ . ?x att$_2$
	val$_2$ . ?x att$_3$ val$_3$\}}), 
while a chain SQL searches for a multi-variable join via indirections
(\struct{SELECT ?x WHERE \{?x att$_1$ ?y . ?y att$_2$ ?z .
	?z att$_3$ val$_1$\}}).
For conversations, this entails the following: in the star model, the entity
in $q^0$ is always assumed to be the entity in all subsequent utterances
(like \struct{The Last Unicorn}). The best
predicate is disambiguated via a search around such $E(q^0)$ using
similarities of word2vec embeddings of Wikidata phrases and non-entity words
in $q^0$. The corresponding missing argument from the triple is returned as 
the answer.
In the chain model of a conversation, the previous answer $a^{t-1}$ is always
taken as the reference entity at turn $t$, instead of $E(q^0)$.
Predicate selection and answer detection are done analogously as in the star
model.

\textbf{No frontiers.} We also investigated whether the idea of
a frontier
node in itself was necessary, by defining an alternative configuration where
we \textit{optimize an answer-scoring objective directly}.
The same three signals of question matching, context proximity,
and KG priors were aggregated
(Eqs.~\ref{eq:match}, \ref{eq:prox}, and \ref{eq:prior}), and
the Fagin's Threshold Algorithm was again applied for obtaining the top-$r$ list.
However, these top-$r$ returned nodes are now directly the answers.
The process used translates to a \textit{branch-and-bound strategy} for
iteratively exploring the neighborhood of
the initial context ($E(q^0)$, $a^0$, and their interconnections) as follows,
without explicitly materializing a context subgraph.
The $2$-hop neighborhood ($2$-hop as we now directly score
for an answer, without finding a frontier first) of each node in the context
at a given turn is scored on its likelihood of being an answer, in
a breadth-first manner. The first computed score defines a lower bound on 
the node being a potential answer, that is updated as better candidates are
found. If a node's answer score is lower than the lower bound so far, it
is not expanded further (its neighborhood is not explored anymore). We keep
exploring the $2$-hop neighborhood of the context iteratively until we do not
find any node in $K$ better than the current best answer. 

\textbf{End-to-end neural model.} We compared our results with
D2A (Dialog-to-Action)~\cite{guo2018dialog}, the state-of-the-art
end-to-end neural model for conversational KG-QA. Since \convex is
an unsupervised method that does not rely on training data, we used the D2A model
pre-trained on the large CSQA benchmark~\cite{saha2018complex}. D2A manages
dialogue memory using a generative model based on a flexible grammar.

\textbf{Question completion.} An interesting question to ask at this stage 
is whether an attempt towards completing the follow-up utterances is worthwhile. 
While a direct adaptation of a method like~\cite{kumar2017incomplete} is
infeasible due to absence of training pairs
and the need for \textit{rewriting } as opposed
to plain \textit{completion}, we investigate certain reasonable alternatives:
(i) when
$q^t$ is concatenated with keywords (all nouns and verbs) from $q^0$;
(ii) when $q^t$ is concatenated with $a^0$;
(iii) when $q^t$ is concatenated with keywords from $q^{t-1}$; and,
(iv) with $a^{i-1}$. These variants are then passed through the stand-alone
KG-QA system. Fortunately, the state-of-the-art system QAnswer is totally
syntax-agnostic, and searches the KG with all question cue words to formulate
an optimal SPARQL query whose components best \textit{cover}
the mapped KG items. This syntax-independent approach was vital as it would be
futile to massage the ``completed'' questions above into grammatically
correct forms. Platypus, on the other hand, is totally dependent on an accurate
dependency parse of the input utterance, and hence is unsuitable for
plugging in these question completion strategies.

\textbf{Metrics.} Since most questions in \convquestions had exactly one
or at most a few correct answers, we used
the standard metrics of
Precision at the top rank (P@1), Mean Reciprocal Rank (MRR), and Hit@5 metrics.
The last measures the fraction of times a correct answer was retrieved within
the top-$5$ positions.

\subsection{Configuration}
\label{subsec:config}

\textbf{Dataset.} We evaluate \convex and other baselines on \convquestions.
A random $20\%$ of the $11k$ conversations was held out for tuning model
parameters, and the remaining $80\%$ was used for testing. Care was taken
that this development set was generated from a separate set of
seed conversations ($70$ out of the original $350$) so as to preclude
possibilities
of ``leakage'' on to the test set.

\textbf{Initialization.} We use Wikidata
(\url{www.wikidata.org}) as our underlying KG,
and use the complete RDF dump in NTriples format from 15 April 2019
(\url{http://bit.ly/2QhsSDC}, $\simeq 1.3$ TB uncompressed).
Identifier triples like those containing predicates like
Freebase ID, IMDb ID, etc. were excluded. 
We used indexing with HDT (\url{www.rdfhdt.org/}) that enables much faster
lookups. The Python library NetworkX
(\url{https://networkx.github.io/}) was used for graph processing.
TagMe was used for NERD, and word2vec embeddings were
obtained via the \textit{spaCy} package.
Stanford CoreNLP~\cite{manning2014stanford} was used for POS
tagging to extract nouns and verbs for question completion.
The ideal number of frontier nodes, $r$, was
found to be \textbf{three} by tuning on the dev set.

\section{Results and Insights}
\label{sec:results}

\subsection{Key findings}
\label{subsec:key}

\begin{table*} [t] \small
	\newcolumntype{G}{>{\columncolor [gray] {0.90}}c}
	\resizebox{\textwidth}{!}{
	\begin{tabular}{l G G G c c c G G G c c c G G G}
		\toprule
		\textbf{Domain}									&	\multicolumn{3}{G}{\textbf{Movies}}												&	\multicolumn{3}{c}{\textbf{TV Series}}											& 	\multicolumn{3}{G}{\textbf{Music}}												& 	\multicolumn{3}{c}{\textbf{Books}}												& 	\multicolumn{3}{G}{\textbf{Soccer}}												\\ \midrule
		\textbf{Method}									&	\textbf{P@1}			&	\textbf{MRR}			&	\textbf{Hit@5}			&	\textbf{P@1}			&	\textbf{MRR}			&	\textbf{Hit@5}			&	\textbf{P@1}			&	\textbf{MRR}			&	\textbf{Hit@5}			&	\textbf{P@1}			&	\textbf{MRR}			&	\textbf{Hit@5}			&	\textbf{P@1}			&	\textbf{MRR}			&	\textbf{Hit@5}			\\ \toprule
		\textbf{QAnswer}~\cite{diefenbach2019qanswer}	&	$0.032$					&	$0.032$					&	$0.032$					&	$0.064$					&	$0.064$					&	$0.064$					&	$0.020$					&	$0.020$					&	$0.020$					&	$0.011$					&	$0.011$					&	$0.011$					&	$0.020$					&	$0.020$					&	$0.020$					\\
		\textbf{QAnswer + \convex}						&	$\boldsymbol{0.222}$*	&	$\boldsymbol{0.264}$*	&	$\boldsymbol{0.311}$*	&	$\boldsymbol{0.136}$*	&	$\boldsymbol{0.172}$*	&	$\boldsymbol{0.214}$*	&	$0.168$					&	$\boldsymbol{0.197}$*	&	$\boldsymbol{0.232}$*	&	$0.177$					&	$\boldsymbol{0.213}$*	&	$\boldsymbol{0.252}$*	&	$\boldsymbol{0.179}$*	&	$\boldsymbol{0.221}$*	&	$\boldsymbol{0.265}$*	\\
		\textbf{QAnswer + Star}							&	$0.201$					&	$0.201$					&	$0.201$					&	$0.132$					&	$0.132$					&	$0.132$					&	$\boldsymbol{0.183}$	&	$0.183$					&	$0.183$					&	$\boldsymbol{0.199}$	&	$0.199$					&	$0.199$					&	$0.170$					&	$0.170$					&	$0.170$					\\
		\textbf{QAnswer + Chain}						&	$0.077$					&	$0.077$					&	$0.077$					&	$0.028$					&	$0.028$					&	$0.028$					&	$0.056$					&	$0.056$					&	$0.056$					&	$0.034$					&	$0.034$					&	$0.034$					&	$0.044$					&	$0.044$					&	$0.044$					\\ \midrule
		\textbf{Platypus}~\cite{tanon2018demoing}		&	$0.000$					&	$0.000$					&	$0.000$					&	$0.000$					&	$0.000$					&	$0.000$					&	$0.005$					&	$0.005$					&	$0.005$					&	$0.002$					&	$0.002$					&	$0.002$					&	$0.004$					&	$0.004$					&	$0.004$					\\
		\textbf{Platypus + \convex}						&	$\boldsymbol{0.218}$*	&	$\boldsymbol{0.255}$*	&	$\boldsymbol{0.295}$*	&	$0.124$					&	$\boldsymbol{0.153}$*	&	$\boldsymbol{0.189}$*	&	$0.167$					&	$\boldsymbol{0.197}$*	&	$\boldsymbol{0.233}$*	&	$0.180$					&	$\boldsymbol{0.216}$*	&	$\boldsymbol{0.256}$*	&	$\boldsymbol{0.179}$*	&	$\boldsymbol{0.222}$*	&	$\boldsymbol{0.269}$*	\\
		\textbf{Platypus + Star}						&	$0.201$					&	$0.201$					&	$0.201$					&	$\boldsymbol{0.132}$	&	$0.132$					&	$0.132$					&	$\boldsymbol{0.183}$	&	$0.183$					&	$0.183$					&	$\boldsymbol{0.199}$	&	$0.199$					&	$0.199$					&	$0.170$					&	$0.170$					&	$0.170$					\\
		\textbf{Platypus + Chain}						&	$0.047$					&	$0.047$					&	$0.047$					&	$0.000$					&	$0.000$					&	$0.000$					&	$0.028$					&	$0.028$					&	$0.028$					&	$0.028$					&	$0.028$					&	$0.028$					&	$0.015$					&	$0.015$					&	$0.015$					\\ \midrule
		\textbf{Naive}									&	$0.016$					&	$0.016$					&	$0.016$					&	$0.020$					&	$0.020$					&	$0.020$					&	$0.021$					&	$0.021$					&	$0.021$					&	$0.007$					&	$0.007$					&	$0.007$					&	$0.016$					&	$0.016$					&	$0.016$					\\
		\textbf{Naive + \convex}						&	$\boldsymbol{0.212}$*	&	$\boldsymbol{0.252}$*	&	$\boldsymbol{0.296}$*	&	$0.121$					&	$\boldsymbol{0.149}$*	&	$\boldsymbol{0.185}$*	&	$0.164$					&	$\boldsymbol{0.194}$*	&	$\boldsymbol{0.229}$*	&	$0.176$					&	$\boldsymbol{0.210}$*	&	$\boldsymbol{0.248}$*	&	$\boldsymbol{0.161}$*	&	$\boldsymbol{0.201}$*	&	$\boldsymbol{0.245}$*	\\
		\textbf{Naive + Star}							&	$0.205$					&	$0.205$					&	$0.205$					&	$\boldsymbol{0.129}$	&	$0.129$					&	$0.129$					&	$\boldsymbol{0.185}$	&	$0.185$					&	$0.185$					&	$\boldsymbol{0.205}$	&	$0.205$					&	$0.205$					&	$0.154$					&	$0.154$					&	$0.154$					\\
		\textbf{Naive + Chain}							&	$0.059$					&	$0.059$					&	$0.059$					&	$0.014$					&	$0.014$					&	$0.014$					&	$0.039$					&	$0.039$					&	$0.039$					&	$0.051$					&	$0.051$					&	$0.051$					&	$0.031$					&	$0.031$					&	$0.031$					\\ \midrule
		\textbf{Oracle + \convex}						&	$\boldsymbol{0.259}$*	&	$\boldsymbol{0.305}$*	&	$\boldsymbol{0.355}$*	&	$0.178$					&	$\boldsymbol{0.218}$*	&	$\boldsymbol{0.269}$*	&	$0.190$					&	$0.237$					&	$\boldsymbol{0.293}$*	&	$0.198$					&	$\boldsymbol{0.246}$*	&	$\boldsymbol{0.303}$*	&	$\boldsymbol{0.188}$*	&	$\boldsymbol{0.234}$*	&	$\boldsymbol{0.284}$*	\\
		\textbf{Oracle + Star}							&	$0.257$					&	$0.257$					&	$0.257$					&	$\boldsymbol{0.194}$	&	$0.194$					&	$0.194$					&	$\boldsymbol{0.241}$	&	$\boldsymbol{0.241}$	&	$0.241$					&	$\boldsymbol{0.241}$	&	$0.241$					&	$0.241$					&	$0.179$					&	$0.179$					&	$0.179$					\\
		\textbf{Oracle + Chain}							&	$0.094$					&	$0.094$					&	$0.094$					&	$0.031$					&	$0.031$					&	$0.031$					&	$0.040$					&	$0.040$					&	$0.040$					&	$0.053$					&	$0.053$					&	$0.053$					&	$0.016$					&	$0.016$					&	$0.016$					\\ \midrule
		\textbf{Oracle + No frontiers}					& 	$0.124$					& 	$0.153$					& 	$0.191$					& 	$0.073$					& 	$0.094$					& 	$0.125$					& 	$0.116$					& 	$0.144$					& 	$0.185$					&	$0.103$					& 	$0.137$					& 	$0.199$					& 	$0.087$					& 	$0.122$					& 	$0.166$					\\ \midrule
		\textbf{D2A}~\cite{guo2018dialog}				&	$0.090$					& 	$0.090$					& 	$0.090$					& 	$0.067$					& 	$0.067$					& 	$0.067$					& 	$0.072$					&	$0.072$					& 	$0.072$					& 	$0.121$					& 	$0.121$					& 	$0.121$					& 	$0.107$					& 	$0.107$					& 	$0.107$					\\ \bottomrule
	\end{tabular}}
	\\ \raggedright The highest value in a group (metric-domain-system triple) is in \textbf{bold}. QAnswer and Platypus return only a top-$1$ answer and not ranked lists,
	and hence have the same P@1, MRR, and Hit@5 values.
	\caption{Our main results on follow-up utterances in \convquestions showing how \convex enables KG-QA enables for conversations, and its comparison with baselines.}
	\label{tab:main-res}
	\vspace*{-0.7cm}
\end{table*}

Table~\ref{tab:main-res} lists main results, where all configurations \textit{are run on the follow-up utterances} in the \convquestions test
($8,960$ conversations; $35,840$ questions). An asterisk (*) indicates statistical significance of \convex-enabled systems over
the strongest baseline in the group, under the $2$-tailed paired $t$-test at $p < 0.05$ level. We make the following key observations.

\textbf{\convex enables stand-alone systems.} The state-of-the-art QAnswer~\cite{diefenbach2019qanswer} scores only about $0.011 - 0.064$
(since it produces sets and not ranked lists, all metric values are identical) on its own on the incomplete utterances, which it is clearly
not capable of addressing. When \convex is applied, its performance jumps significantly to $0.172 - 0.264$ ) (MRR) across the domains. We have exactly the
same trends with the Platypus system. The naive strategy with direct entity and predicate linking performs hopelessly in absence of explicit cues, but with
\convex we again see noticeable improvements, brought in by a relevant context graph and its iterative expansion. In the Oracle method, $a^0$ is known,
and hence a row by itself is not meaningful. However, contrasting Oracle+\convex with other \phrase{+\convex} methods, we see that there is 
significant room for improvement that can be achieved by answering $q^0$ correctly.

\textbf{Star and chain models of conversations fall short.} For every configuration, 
we see the across-the-board superiority of \convex-boosted
methods over star- and chain-models (often over $100\%$ gains). This clearly indicates that while these are intuitive ways of modeling human conversation (as seen in the often
respectable values that these achieve), they are insufficient and oversimplified. Evidently, real humans rather prefer the middle path: sometimes hovering
around the initial entity, sometimes drifting in a chain of answers. A core component of \convex that we can attribute this pattern to, is the turn-based
weighting of answer and context proximity that prefers entities in the first and the last turns.
``QAnswer + Star'' and ``Platypus + Star'' achieve the same values as they both operate around the same $q^0$ entity detected by TagMe.


\textbf{\convex generalizes across domains.} In Table~\ref{tab:main-res}, we
also note that the performance of \convex stretches across all five domains (even though the nature of 
questions in each of these domains have their own peculiarities), showing the potential of 
of our unsupervised approach in new domains with little training resources, or to deal with cold starts in enterprise applications. While we did tune
hyperparameters individually for each domain, there were surprisingly little variation across them ($h^F_1 \simeq 0.5-0.6, h^F_2 \simeq 0.3-0.4,
h^F_3 \simeq 0.1, h^A_1 \simeq 0.8-0.9, h^A_2 \simeq 0.1-0.2$).

\textbf{Frontiers help.} We applied our frontier-less approach over the oracle annotations for $q^0$, and in the row marked
``Oracle + No frontiers'' in Table~\ref{tab:main-res}, we find that this results in degraded performance.
We thus claim that locating frontiers is
an essential step before answer detection. The primary reason behind
this is that answers only have low direct matching similarity to the question, making a $2$-stage approach worthwhile.
Also, exploring a $2$-hop neighborhood was generally found to suffice: nodes further away from the initial context rarely manage
to ``win'', due to the proximity score component quickly falling off as KG-hops increase.

\textbf{Pre-trained models do not suffice.} D2A produces a single answer for every utterance,
which is why the three metrics are equal.
From the D2A row in Table~\ref{tab:main-res},
we observe that pre-trained neural models do not work well off-the-shelf on \convquestions
(when compared to the \convex-enabled QAnswer row, for example).
This is mostly due to the restrictive patterns in the CSQA dataset, owing to
its semi-synthetic mode of creation. A direct comparison, though, is not fair, as \convex
is an enabler method for a stand-alone KG-QA system, while D2A is an end-to-end model.
Nevertheless, the main classes of errors come from: (i) a predicate necessary in \convquestions
that is absent in CSQA (D2A cannot answer temporal questions like
\utterance{In what year was Ender's game written?} as such relations are absent in CSQA);
(ii) D2A cannot generate $2$-hop KG triple patterns;
(iii) D2A cannot resolve long-term co-references in questions (pronouns only come from
the last turn in CSQA, but not in \convquestions);
(iv) In CSQA, co-references are almost always
indicated as ``it'' or ``that one''. But since \convquestions is completely user-generated,
we have more challenging cases with ``this book'', ``the author'', ``that year'', and so on.

\begin{table} [t] 
	\centering
	\resizebox*{\columnwidth}{!}{
	\begin{tabular}{l c c c c c}
		\toprule
		\textbf{Method}					&	\textbf{Movies}			&	\textbf{TV}				&	\textbf{Music}			&	\textbf{Books}			&	\textbf{Soccer} 		\\	\toprule
		QAnswer + \convex				&	$\boldsymbol{0.264}$*	&	$\boldsymbol{0.172}$*	&	$\boldsymbol{0.197}$*	&	$\boldsymbol{0.213}$*	&	$\boldsymbol{0.221}$* 	\\ \midrule
		QAnswer + $q^0$ keywords		&	$0.071$					&	$0.052$					&	$0.084$					&	$0.039$					&	$0.075$ 				\\
		QAnswer + $a^0$					&	$0.077$					&	$0.054$					&	$0.048$					&	$0.096$					&	$0.045$ 				\\
		QAnswer + $q^{i-1}$ keywords	&	$0.050$					&	$0.045$					&	$0.045$					&	$0.025$					&	$0.046$ 				\\
		QAnswer + $a^{i-1}$				&	$0.109$					&	$0.079$					&	$0.093$					&	$0.064$					&	$0.070$ 				\\ \bottomrule
	\end{tabular}}
	\caption{Comparison with question completion strategies (MRR). The highest value in a column is in \textbf{bold}.}
	\label{tab:qcomp}
	\vspace*{-0.3cm}
\end{table}

\textbf{\convex outperforms question completion methods.} Comparison with question completion methods are presented in
Table~\ref{tab:qcomp}. Clear trends show that while these strategies generally perform better than stand-alone systems (contrasting QAnswer
with Table~\ref{tab:main-res}, for, say, Movies, we see $0.050-0.109$ vs. $0.032$ on MRR previously), use of \convex results in
higher improvement ($0.264$ MRR on Movies).
This implies that question completion
is hardly worthwhile in this setup when the KG structure already reveals a great deal about the underlying user intents left implicit
in follow-up utterances.

\subsection{Analysis}
\label{subsec:ana}

\begin{table} [t] \small
	\centering
		\begin{tabular}{c c c c c c}
			\toprule
			\textbf{Turn}	&	\textbf{Movies}	&	\textbf{TV}	&	\textbf{Music}	&	\textbf{Books}	&	\textbf{Soccer}	\\ \toprule
			\textbf{1}		&	$0.375$			&	$0.393$		&	$0.080$			&	$0.446$			&	$0.357$			\\
			\textbf{2}		&	$0.375$			&	$0.250$		&	$0.214$			&	$0.281$			&	$0.188$			\\
			\textbf{3}		&	$0.161$			&	$0.205$		&	$0.124$			&	$0.435$			&	$0.304$			\\
			\textbf{4}		&	$0.325$			&	$0.214$		&	$0.044$			&	$0.375$			&	$0.137$			\\ \bottomrule
	\end{tabular}
	\caption{Performance of \convex over turns (MRR).}
	\label{tab:turns}
	\vspace*{-0.3cm}
\end{table}


\textbf{\convex maintains its performance over turns.} One of the most promising results of this zoomed-in analysis
is that the MRR for \convex (measured via its combination with the Oracle, to decouple the effect of the QA system)
does not diminish over turns. This shows particular robustness of our graph-based method: while we may produce
several wrong results during the session of the conversation, we are not bogged down by any single mistake, as the
context graph retains several scored candidates within itself, guarding against ``near misses''. This is in stark contrast
to the chain model, where it is exclusively dependent on $a^{t-1}$.

\textbf{Error analysis.} \convex has two main steps in its pipeline: context expansion, and answer ranking. Analogously,
there are two main cases of error: when the answer is not pulled in when $X^t$ is expanded at the frontiers (incorrect
frontier scoring), or when the answer is there in $X^t_+$ but is not ranked at the top. These numbers are shown in 
Table~\ref{tab:error}. We find that there is significant scope for improvement for frontier expansion, as $80-90\%$ errors
lie in this bag. It is however, heartening to see that no particular turn is singularly affected. This calls for more
informed frontier scoring than our current strategy. Answer ranking can be improved with
better ways of aggregating the two proximity signals. Table~\ref{tab:anec} lists anecdotal examples of success cases
with \convex.

\begin{table} [t] \small
	\centering
	\resizebox*{\columnwidth}{!}{
	\begin{tabular}{l c c c c}
		\toprule
		\textbf{Scenario} 							&	\textbf{Turn 1}	&	\textbf{Turn 2}	&	\textbf{Turn 3}	&	\textbf{Turn 4} \\ \toprule
		Ans. not in expanded graph					& 	$87.1$			& 	$79.8$			&	$89.2$			&	$89.6$			\\
		Ans. in expanded graph but not in top-$1$	& 	$12.9$			& 	$20.2$			&	$10.8$			&	$10.4$			\\ \bottomrule
	\end{tabular}}
	\caption{Error analysis (percentages of total errors).}
	\label{tab:error}
	\vspace*{-0.3cm}
\end{table}

\begin{table} [t] \small
	\centering
	\resizebox*{\columnwidth}{!}{
	\begin{tabular}{l}
		\toprule
		\textbf{Utterance:} \utterance{What was the name of the director?} (Movies, Turn 4)									\\
		\textbf{Intent:} \utterance{Who was the director of the movie Breakfast at Tiffany's?}								\\ \midrule
		\textbf{Utterance:} \utterance{What about Mr Morningstar?} (TV Series, Turn 2) 										\\
		\textbf{Intent:} \utterance{Which actor plays the role of Mr Morningstar in the TV series Lucifer?} 				\\ \midrule
		\textbf{Utterance:} \utterance{What record label put out the album?} (Music, Turn 3) 								\\
		\textbf{Intent:} \utterance{What is the name of the record label of the album Cosmic Thing?} 						\\ \midrule
		\textbf{Utterance:} \utterance{written in country?} (Books, Turn 4) 												\\
		\textbf{Intent:} \utterance{In which country was the book ``The Body in the Library'' by Agatha Christie written?} 	\\ \midrule
		\textbf{Utterance:} \utterance{Who won the World Cup that year?} (Soccer, Turn 4) 									\\
		\textbf{Intent:} \utterance{Which national team won the 2010 FIFA World Cup?} 										\\ \bottomrule
	\end{tabular}}
	\caption{Representative examples where Oracle + \convex produced the best answer at
		the top-1, but neither Oracle + Star, nor Oracle + Chain could.}
	\label{tab:anec}
	\vspace*{-0.3cm}
\end{table}

\section{Related Work}
\label{sec:related}

\textbf{Question answering over KGs.} Starting with early approaches
in 2012-'13~\cite{unger2012template,yahya2013robust,berant2013semantic}, based
on parsing questions via handcoded templates and grammars, KG-QA already has
a rich body of literature. While templates continued to be a strong line of work
due to its focus on interpretability and
generalizability~\cite{bast2015more,abujabal:17,abujabal2018never,
	diefenbach2019qanswer,tanon2018demoing},
a parallel thread has focused on neural methods driven by
performance
gains~\cite{huang2019knowledge,lukovnikov2017neural,sawant2019neural}. 
Newer trends
include shifts towards more
complex questions~\cite{luo2018knowledge,talmor2018web,lu2019answering},
and fusion of 
knowledge graphs
and text~\cite{sawant2019neural,sun2018open}. However, none of these approaches
can deal with incomplete questions in a conversational setting.

\textbf{Conversational question answering.} Saha et al.~\cite{saha2018complex}
introduce the paradigm of sequential question answering over KGs, and create
a large benchmark CSQA for the task, along
with a baseline with memory networks.
Guo et al.~\cite{guo2018dialog} propose D2A,
an end-to-end technique for conversational KG-QA ,
that introduces dialog memory management 
for inferring the logical form of
current utterances. 
While our goal is rather to build a conversation enabler method, we still compare with,
and outperform the CSQA-trained D2A model on \convquestions.

\textit{Question completion}
approaches~\cite{kumar2017incomplete,raghu2015statistical,ren2018conversational}
target
this setting by attempting to create full-fledged
interrogatives from partial utterances while being independent of the answering
resource, but suffer in situations without training pairs and with ad hoc styles.
Nevertheless, we try to compare with this line of thought, and show that
such completion may not be necessary if the underlying KG can be properly
exploited.

Iyyer et al.~\cite{iyyer2017search} initiate the direction of
sequential QA over \textit{tables} using dynamic neural semantic parsing trained via
weakly supervised reward-guided search, and evaluate by decomposing a
previous benchmark of complex questions~\cite{liang2015compositional} to create
sequential utterances. However, such table-cell search methods cannot scale
to real-world, large-scale curated KGs. 

QBLink~\cite{elgohary2018dataset},
CoQA~\cite{reddy2018coqa},
and ShARC~\cite{saeidi2018interpretation} are recent benchmarks aimed at driving
conversational QA \textit{over text}, and the allied paradigm in text
comprehension on interactive QA~\cite{li2017context}.
Hixon et al.~\cite{hixon2015learning} try to learn concept knowledge graphs from
conversational dialogues over science questions, but such KGs are
fundamentally different from curated ones like Wikidata with millions of facts.
\section{Conclusion}
\label{sec:conclusion}

%
%
%
%
Through \convex, we showed how judicious graph expansion strategies with
informed look-ahead, can help stand-alone KG-QA systems cope with some of
the challenges posed by incomplete and ad hoc follow-up questions in fact-centric
conversations. \convex is completely unsupervised, and thus can be readily
applied to new domains, or be deployed in enterprise setups with little or no
training data. Further, being a graph-based method, each turn and the
associated expansions can be easily visualized, resulting in interpretable
evidence
for answer derivation: an unsolved concern for many neural methods for QA. Nevertheless, 
\convex is just a first step 
towards solving
the challenge of conversational KG-QA. We believe that the \convquestions
benchmark, reflecting real user behavior, can play a key role 
in driving further progress.

\textbf{Acknowledgements.} We sincerely thank Daya Guo (Sun Yat-sen University)
for his help in executing D2A on \convquestions.

\bibliographystyle{ACM-Reference-Format}
\bibliography{convex}

\end{document}